\documentclass[aps,PRB,twocolumn,superscriptaddress,preprintnumbers]{revtex4-2}
\usepackage[T1]{fontenc}
\usepackage{times}
\usepackage{graphicx,color}
\usepackage{amsfonts,amsmath,amssymb,amsbsy}
\usepackage[colorlinks=true, linkcolor=blue, citecolor=blue, urlcolor=blue]{hyperref}
\usepackage{float}

\def\beginTable{\begin{table}[h]}
\def\endTable{\end{table}}

\def\beginWide{\begin{widetext}}
\def\endWide{\end{widetext}}
\begin{document}

\title{Fourth-order and six-order nonlinear spin current  rectifier in  three-dimensional  $h$-wave and $j$-wave odd-parity magnets}
\author{Motohiko Ezawa}
\affiliation{Department of Applied Physics, The University of Tokyo, 7-3-1 Hongo, Tokyo
113-8656, Japan}

\begin{abstract}
Higher-order symmetric $X$-wave magnets consist of two groups. One includes $d$-wave, $g$-wave and $i$-wave altermagnets, while the other includes $p$-wave and $f$-wave odd-parity magnets. Recently, the possibility of $h$-wave magnets has been discussed. Motivated by this development, we systematically construct an $X$-wave magnet with $\left( N_{X}+1\right) $ nodes in three dimensions from an $X$-wave magnet with $N_{X}$ nodes in two dimensions by means of a dimensional extension, where $N_X=1,2,3,4,6$ for $X=p,d,f,g,i$, respectively. Based on this method, we predict $j$-wave magnets in three dimensions. Then, we argue how to identify each of these $X$-wave magnets experimentally. We show that the $X$-wave magnet is completely identified by measuring the nonlinear spin currents. In particular, we predict that there are no spin currents other than the fourth-order ones such as $\sigma _{\text{spin}}^{x^{3}y;z}$ in $h$-wave odd-parity magnets in three dimensions and the sixth-order ones such as $\sigma _{\text{spin}}^{x^{5}y;z}$ in $j$-wave odd-parity magnets in three dimensions. They function as spin-current  rectifier  because the spin current exhibits unidirectional flow independent of the applied electric field.
\end{abstract}

\date{\today }
\maketitle

\section{Introduction}

Altermagnets\cite{SmejX,SmejX2,Naka,Gonza,NakaB,Bose,NakaRev} are one of the
most exciting fields in condensed matter physics. It will enable us to
realize an ultrafast and ultradence spintronic memory due to a
characteristic spin-split band structure and zero net magnetization.
Altermagnets are characterized by breaking time-reversal symmetry and
preserving inversion symmetry. In addition, $p$-wave magnets\cite%
{Hayami,pwave,He,Comin,Yamada,HZhou,Brek,EzawaPNeel,Chak,Atasi,PEdel} and $f$%
-wave magnets\cite{pwave,He,BitanRoy,GI,NLSeebeck} also attract attention
although they are not altermagnets. They are characterized by preserving
time-reversal symmetry and breaking inversion symmetry. In this sense, they
are odd-parity magnets. These altermagnets and odd-parity magnets are
summarized in terms of $X$-wave magnets\cite{Planar,MTJ,APEX} with $%
X=p,d,f,g,i$. They are described by the band-splitting functions $f_{X}(k)$\
whose number of nodes are $N_{X}=1,2,3,4,6$, respectively. See Table \ref%
{TabXwave}. The $X$-wave magnet with $N_{X}=5$ is absent in two dimensions
(2D) because the five-fold rotational symmetry is prohibited in crystal.
Recently, $h$-wave magnets with $N_{X}=5$ were proposed\cite{YuH} in three
dimensions (3D) but their physical properties are yet to be explored.
Indeed, FeSe has been argued to be a $h$-wave magnet based on the
density-functional theory\cite{YuH}.

Nonlinear spin current generation is an important topic in spintronics\cite%
{Hamamoto,Kameda,Hayami22B,Hayami24B}. It is prominent that only a certain
order of nonlinear spin currents are generated in an $X$-wave magnet\cite%
{GI,NLSeebeck}. The order of nonlinearity of spin currents is determined by
the number of nodes $N_{X}$ of the $X$-wave magnet. For instance, only the
second-order nonlinear spin currents are generated in $f$-wave magnets\cite%
{GI}. In a similar way, only the third-order (fifth-order) nonlinear spin
currents are generated in $g$-wave ($i$-wave) altermagnets. See Table \ref%
{TabXwave}.

In this paper, we propose a method to systematically construct an $X$-wave
magnet with $\left( N_{X}+1\right) $ nodes in 3D from an $X$-wave magnet
with $N_{X}$ nodes in 2D by means of a dimensional extension. In this
framework, it is naturally understood that $h$-wave magnets are 3D extension
of $g$-wave altermagnets. Similarly, we predict $j$-wave magnets in 3D by a
dimensional extension of $i$-wave altermagnets. A candidate crystal
structure is a triangular prism lattice.  We represent group-theoretical
analysis for $X$-wave magnets.  There are no other $X$-wave magnets in 3D
which can be constructed by this method because there are no $X$-wave
magnets with $N_{X}>6$ in 2D. Then, we show that the $X$-wave magnet is
completely identified by measuring the nonlinear spin conductivities. In
particular, we predict that only the fourth-order nonlinear spin currents
such as $\sigma _{\text{spin}}^{x^{3}y;z}$ emerge in $h$-wave odd-parity
magnets in 3D and only the sixth-order ones such as $\sigma _{\text{spin}%
}^{x^{5}y;z}$ in $j$-wave odd-parity magnets in 3D. See Table \ref{TabXwave}%
. They function as spin-current  rectifier  because the spin current
exhibits unidirectional flow independent of the applied electric field.

\beginTable%
\begin{tabular}{|c|c|c|c|c|c|c|c|}
\hline
$X$ & $p$ & $d$ & $f$ & $g$ & $h$ & $i$ & $j$ \\ \hline
$N_{X}$ & 1 & 2 & 3 & 4 & 5 & 6 & 7 \\ \hline
$\ell $ &  & linear & 2nd NL & 3rd NL & 4th NL & 5th NL & 6th NL \\ \hline
2D & 0 & $\sigma _{\text{spin}}^{y;x}$ & $\sigma _{\text{spin}}^{x^{2};y}$ & 
$\sigma _{\text{spin}}^{y^{3};x}$ & $\times $ & $\sigma _{\text{spin}%
}^{y^{5};x}$ & $\times $ \\ \hline
3D & 0 & $\sigma _{\text{spin}}^{y;x}$ & $\sigma _{\text{spin}}^{zy;x}$ & $%
\sigma _{\text{spin}}^{x^{3};z}$ & $\sigma _{\text{spin}}^{x^{3}y;z}$ & $%
\sigma _{\text{spin}}^{y^{4}x;x}$ & $\sigma _{\text{spin}}^{x^{5}y;z}$ \\ 
\hline
\end{tabular}%
\caption{$X$-wave magnets and nonlinear spin conductivity. The $X$-wave magnet is
characterized by the number of nodes $N_{X}$, where only the $\ell $-th
order nonlinear spin currents are generated. Typical spin conductivities are
listed in 2D and 3D.} \label{TabXwave} \endTable

This paper is organized as follows. Section II summarizes known $X$-wave
magnets and proposes three-dimensional $h$-wave and $j$-wave magnets based
on the layer-construction method. Section III discusses the point groups,
magnetic point groups, and irreducible representations relevant to $h$-wave
and $j$-wave magnets. In Section IV, we determine the Fermi surfaces of $h$%
-wave and $j$-wave magnets and present the corresponding tight-binding
models. Section V evaluates the nonlinear spin current in these systems.
Section VI is devoted to discussions. Appendix A provides point groups,
magnetic point groups, and irreducible representations for all $X$-wave
magnets.

\section{$X$-wave magnets}

The electromagnetic property of the $X$-wave magnet is characterized by the
band splitting depending on the spin. The simplest model is the two-band
Hamiltonian\cite{GI,Planar,MTJ,APEX,NLSeebeck} given by%
\begin{equation}
H\left( \mathbf{k}\right) =\frac{\hbar ^{2}k^{2}}{2m}+Jf_{X}\left( \mathbf{k}%
\right) \sigma _{z},  \label{Model}
\end{equation}%
where the first term represents the kinetic term of electrons, while the
second term represents the\ band splitting described by the function $%
f_{X}\left( \mathbf{k}\right) $\ with the coupling constant $J$\ and the
Pauli matrix $\sigma _{z}$. The energy is given by 
\begin{equation}
\varepsilon _{s}\left( \mathbf{k}\right) =\frac{\hbar ^{2}k^{2}}{2m}%
+sJf_{X}\left( \mathbf{k}\right)
\end{equation}
with $s=\pm 1$, where $s=1$ for up spin and $s=-1$ for down spin.

\begin{figure}[t]
\centerline{\includegraphics[width=0.48\textwidth]{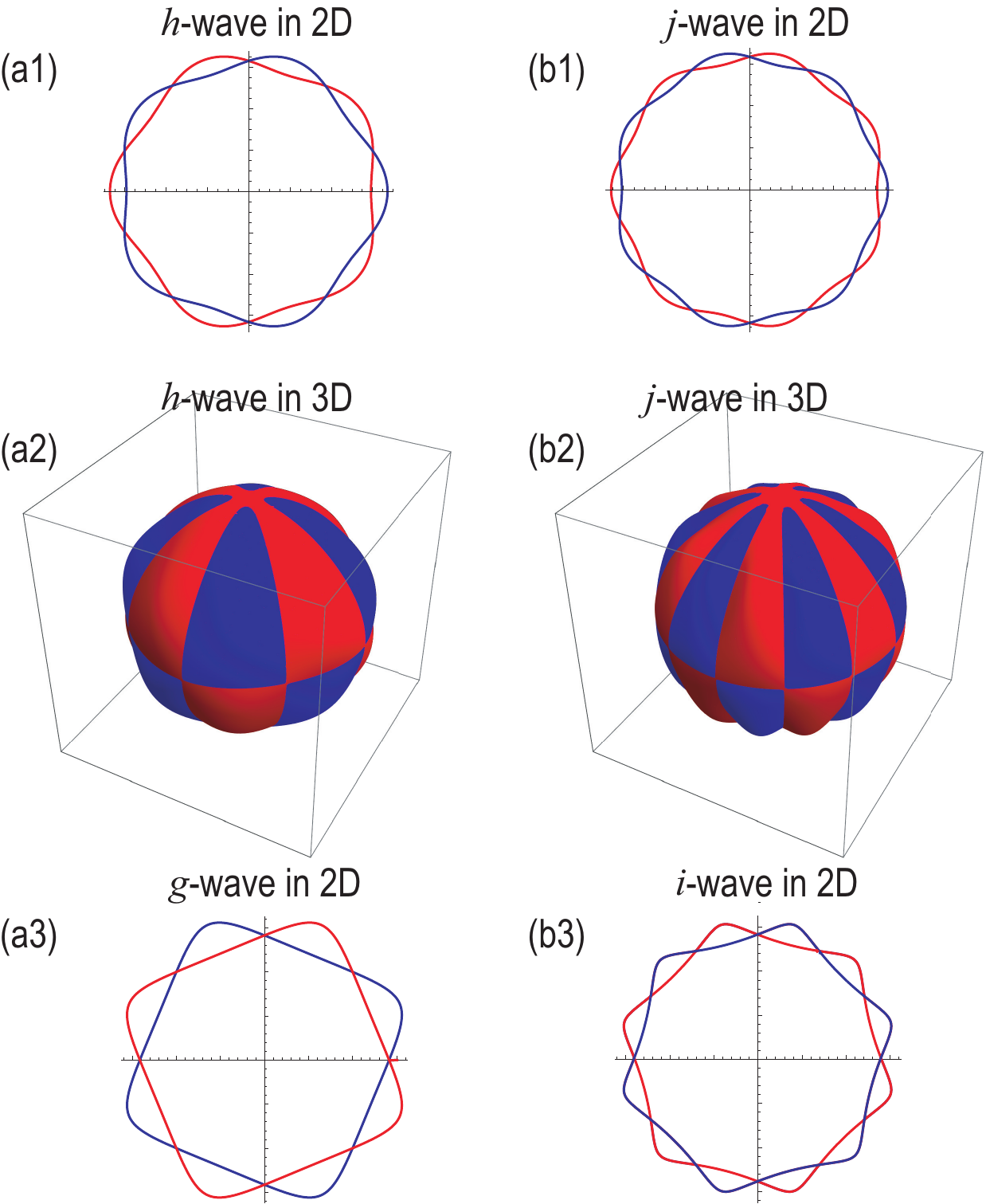}}
\caption{Illustration of the Fermi surface. (a1) $h$-wave magnets in 2D.
(b1) $j$-wave magnets in 2D. (a2) $h$-wave magnets in 3D. (b2) $j$-wave
magnets in 3D. (a3) $g$-wave altermagnet in 3D. (b3) $i$-wave altermagnet in
2D. There is no lattice realization of $h$-wave ($j$-wave) magnets in 2D due
to the five-fold (seven-fold) rotational symmetry. The $h$-wave ($j$-wave)
magnet in 3D is constructed as a dimensional extension of the $g$-wave ($i$%
-wave) altermagnet in 2D. }
\label{FigIllust}
\end{figure}

In 2D, with the use of the lattice constant $a$, they are given by\cite%
{SmejX,SmejX2,GI,Planar,MTJ},%
\begin{align}
f_{s}^{2\text{D}}\left( \mathbf{k}\right) & =1, \\
f_{p}^{2\text{D}}\left( \mathbf{k}\right) & =ak_{x}=ak\cos \phi ,
\label{f2p} \\
f_{d}^{2\text{D}}\left( \mathbf{k}\right) & =2a^{2}k_{x}k_{y}=a^{2}k^{2}\sin
2\phi ,  \label{f2d} \\
f_{f}^{2\text{D}}\left( \mathbf{k}\right) & =a^{3}k_{x}\left(
k_{x}^{2}-3k_{y}^{2}\right) =a^{3}k^{3}\cos 3\phi , \\
f_{g}^{2\text{D}}\left( \mathbf{k}\right) & =4a^{4}k_{x}k_{y}\left(
k_{x}^{2}-k_{y}^{2}\right) =a^{4}k^{4}\sin 4\phi , \\
f_{i}^{2\text{D}}\left( \mathbf{k}\right) & =2a^{6}k_{x}k_{y}\left(
3k_{x}^{2}-k_{y}^{2}\right) \left( k_{x}^{2}-3k_{y}^{2}\right)
=a^{6}k^{6}\sin 6\phi ,
\end{align}%
and%
\begin{align}
f_{p^{\prime }}^{2\text{D}}\left( \mathbf{k}\right) =& ak_{y}=ak\sin \phi ,
\\
f_{d^{\prime }}^{2\text{D}}\left( \mathbf{k}\right) =& a^{2}\left(
k_{x}^{2}-k_{y}^{2}\right) =a^{2}k^{2}\cos 2\phi , \\
f_{f^{\prime }}^{2\text{D}}\left( \mathbf{k}\right) =& a^{3}k_{y}\left(
3k_{x}^{2}-k_{y}^{2}\right) =a^{3}k^{3}\sin 3\phi , \\
f_{g^{\prime }}^{2\text{D}}\left( \mathbf{k}\right) =& a^{4}\left(
k_{x}^{4}-6k_{x}^{2}k_{y}^{2}+k_{y}^{4}\right)  \notag \\
=& a^{4}k^{4}\cos 4\phi , \\
f_{i^{\prime }}^{2\text{D}}\left( \mathbf{k}\right) =& 2a^{6}\left(
k_{x}^{2}-k_{y}^{2}\right) \left(
k_{x}^{4}+k_{y}^{4}-14k_{x}^{2}k_{y}^{2}\right)  \notag \\
=& a^{6}k^{6}\cos 6\phi ,
\end{align}%
each of which has $N_{X}$ nodes. In 3D, they are given by 
\begin{align}
f_{d}^{3\text{D}}\left( \mathbf{k}\right) & =a^{2}k_{z}\left(
k_{x}+k_{y}\right) , \\
f_{f}^{3\text{D}}\left( \mathbf{k}\right) &
=2a^{3}k_{x}k_{y}k_{z}=a^{2}k^{2}\cos \theta \sin 2\phi , \\
f_{g}^{3\text{D}}\left( \mathbf{k}\right) & =a^{4}k_{z}k_{x}\left(
k_{x}^{2}-3k_{y}^{2}\right) =a^{3}k^{3}\cos \theta \cos 3\phi , \\
f_{i}^{3\text{D}}\left( \mathbf{k}\right) & =a^{6}\left(
k_{x}^{2}-k_{y}^{2}\right) \left( k_{y}^{2}-k_{z}^{2}\right) \left(
k_{z}^{2}-k_{x}^{2}\right) ,  \label{EqY}
\end{align}%
and%
\begin{align}
f_{d^{\prime }}^{3\text{D}}\left( \mathbf{k}\right) =& a^{2}k_{y}\left(
k_{x}+k_{z}\right) , \\
f_{f^{\prime }}^{3\text{D}}\left( \mathbf{k}\right) =& a^{3}k_{z}\left(
k_{x}^{2}-k_{y}^{2}\right) =a^{2}k^{2}\cos \theta \cos 2\phi , \\
f_{g^{\prime }}^{3\text{D}}\left( \mathbf{k}\right) =& a^{4}k_{z}k_{y}\left(
3k_{x}^{2}-k_{y}^{2}\right) =a^{3}k^{3}\cos \theta \sin 3\phi .
\end{align}%
each of which has $N_{X}$ nodes. There are relations between an $X$-wave
magnet with $N_{X}+1$ in 3D and\ an $X$-wave magnet with $N_{X}$ in 2D such
as \cite{APEX},%
\begin{align}
f_{f}^{3\text{D}}\left( \mathbf{k}\right) =a& k_{z}f_{d}^{2\text{D}}\left( 
\mathbf{k}\right) , \\
f_{g}^{3\text{D}}\left( \mathbf{k}\right) =a& k_{z}f_{f}^{2\text{D}}\left( 
\mathbf{k}\right) .
\end{align}%
We propose to construct $h$-wave magnets and $j$-wave magnets in 3D from the 
$g$-wave and $i$-wave altermagnets in 2D by a dimensional extension as%
\begin{align}
f_{h}^{3\text{D}}\left( \mathbf{k}\right) =a& k_{z}f_{g}^{2\text{D}}\left( 
\mathbf{k}\right) ,  \label{f} \\
f_{j}^{3\text{D}}\left( \mathbf{k}\right) =a& k_{z}f_{i}^{2\text{D}}\left( 
\mathbf{k}\right) ,  \label{ff}
\end{align}%
provided that there are crystal structures compatible with $f_{h}^{3\text{D}%
}\left( \mathbf{k}\right) $ and $f_{j}^{3\text{D}}\left( \mathbf{k}\right) $%
. We note that the $i$-wave altermagnet in 3D is not constructed in this way
because there is no $h$-wave odd-parity magnet in 2D.

\section{Group theoretical analysis}

We classify $X$-wave magnets based on spherical-harmonic functions and
crystallographic point groups as in the case of classification of multipoles%
\cite{Watanabe,Kusunose}. Multipoles are expressed by the spherical-harmonic
function as

\begin{equation}
Q_{\ell m}\equiv -e\sqrt{\frac{4\pi }{2\ell +1}}k^{\ell }Y_{\ell m}\left(
\theta ,\phi \right) .
\end{equation}%
The $f_{X}$ terms correspond to 
\begin{eqnarray}
Q_{\ell m}^{+} &\equiv &\frac{\left( -1\right) ^{m}}{\sqrt{2}}\left( Q_{\ell
m}+Q_{\ell m}^{\ast }\right) , \\
Q_{\ell m}^{-} &\equiv &\frac{\left( -1\right) ^{m}}{i\sqrt{2}}\left(
Q_{\ell m}-Q_{\ell m}^{\ast }\right) .
\end{eqnarray}%
The explicit forms are given by%
\begin{eqnarray}
Q_{\ell m}^{+} &\propto &k^{\ell -m}P_{\ell }^{m}\left( \cos \theta \right) 
\text{Re}\left[ k_{+}^{m}\right] , \\
Q_{\ell m}^{-} &\propto &k^{\ell -m}P_{\ell }^{m}\left( \cos \theta \right) 
\text{Im}\left[ k_{+}^{m}\right] ,
\end{eqnarray}%
where $P_{\ell }^{m}$ is the associated Legendre polynimial and we have used
the relation%
\begin{equation}
Y_{\ell m}\left( \theta ,\phi \right) =\left( -1\right) ^{\frac{\ell
+\left\vert \ell \right\vert }{2}}\sqrt{\frac{2\ell +1}{4\pi }\frac{\left(
\ell -\left\vert m\right\vert \right) !}{\left( \ell +\left\vert
m\right\vert \right) !}}P_{\ell }^{\left\vert m\right\vert }\left( \cos
\theta \right) e^{im\phi }.
\end{equation}

There are relations%
\begin{eqnarray}
Q_{N_{X},N_{X}-1}^{+} &\propto &k_{z}Q_{N_{X}-1,N_{X}-1}^{+}, \\
Q_{N_{X},N_{X}-1}^{-} &\propto &k_{z}Q_{N_{X}-1,N_{X}-1}^{-},
\end{eqnarray}%
and%
\begin{eqnarray}
f_{X}^{2D} &=&Q_{N_{X}N_{X}}^{-},\qquad f_{X^{\prime
}}^{2D}=Q_{N_{X}N_{X}}^{+}, \\
f_{X}^{3D} &=&Q_{N_{X},N_{X}-1}^{+},\qquad f_{X^{\prime
}}^{3D}=Q_{N_{X},N_{X}-1}^{-}.
\end{eqnarray}

We show the classification table for $h$-wave and $j$-wave magnets in terms
of the point group (PG), the magnetic point group (MPG) and the irreducible
representation (IR): See Table \ref{Tab-hj}. First, $h$-wave magnets are
realized in the orthorhombic crystal system, which belongs to the $D_{4h}$
point group. The explicit crystal structures are the perovskite structure,
the Rutile structure, the olivine structure and the Illite structure. On the
other hand, $j$-wave magnets are realized in the hexagonal crystal, which
belongs to the $D_{6h}$ group. The explicit crystal structures are the
Delafossite-type structure (ABO$_{2}$), 1T/2H systems made of
transition-metal dichalcogenides (TMDs), CdI$_{2}$-type, NaCoO$_{2}$ / Na$%
_{x}$CoO$_{2}$ structure and the YbMgGaO$_{4}$ structure. Magnetic point
groups relevant to $X$-wave magnets belong to type-III (black-white)
magnetic point groups, in which the magnetization is reversed by the $N_{X}$
rotation. These groups take forms such as $N/mmm$ and $N^{\prime }/m^{\prime
}m^{\prime }m^{\prime }$, where $N$ or $N^{\prime }$ indicates whether the
magnetization is preserved or reversed under a $2\pi /N$ rotation. Likewise, 
$m$ or $m^{\prime }$ specifies whether the magnetization is preserved or
reversed by the corresponding mirror operation. For the sake of the
completeness, we classify all $X$-wave magnets in Appendix A.

\beginWide

\begin{center}
\beginTable%
\begin{tabular}{|c|c|c|c|c|}
\hline
&  & \text{Spherical harmonic functions} & \text{PG (MPG)} & \text{IR} \\ 
\hline
$f_{h}^{3\text{D}}$ & $Q_{54}^{-}$ & $k_{z}k_{x}k_{y}\left(
k_{x}^{2}-k_{y}^{2}\right) $ & $D_{4h}(4/m^{\prime }m^{\prime }m^{\prime })$
& $A_{1u}$ \\ \hline
$f_{h^{\prime }}^{3\text{D}}$ & $Q_{54}^{+}$ & $k_{z}\left(
k_{x}^{4}-6k_{x}^{2}k_{y}^{2}+k_{y}^{4}\right) $ & $D_{4h}(4/mmm^{\prime })$
& $A_{2u}$ \\ \hline
$f_{j}^{3\text{D}}$ & $Q_{76}^{-}$ & $2k_{z}k_{x}k_{y}\left(
3k_{x}^{2}-k_{y}^{2}\right) \left( k_{x}^{2}-3k_{y}^{2}\right) $ & $%
D_{6h}(6/m^{\prime }m^{\prime }m^{\prime })$ & $A_{2u}$ \\ \hline
$f_{j^{\prime }}^{3\text{D}}$ & $Q_{76}^{+}$ & $k_{z}\left(
k_{x}^{6}-15k_{x}^{4}k_{y}^{2}+15k_{x}^{2}k_{y}^{2}-k_{y}^{6}\right) $ & $%
D_{6h}(6/mmm^{\prime })$ & $A_{1u}$ \\ \hline
\end{tabular}
\caption{Classification table for $h$-wave and $j$-wave magnets in terms of the point group
(PG), the magnetic point group (MPG) and the irreducible representation (IR). }
\label{Tab-hj} \endTable
\end{center}

\endWide

\section{$h$-wave and $j$-wave magnets}

\subsection{$h$-wave magnets in 2D}

The continuum form of the $h$-wave magnet in 2D is described by%
\begin{equation}
f_{h}^{2\text{D}}\left( \mathbf{k}\right) =k^{5}\cos 5\phi .  \label{f2h}
\end{equation}%
The Fermi surface is given by%
\begin{equation}
k_{\text{F}}=\frac{\sqrt{2m\mu }}{\hbar }-sJ\frac{4\mu ^{2}m^{3}}{\hbar ^{6}}%
\cos 5\phi
\end{equation}%
at $\varepsilon =\mu $ up to the first order in $J$, and shown in Fig.\ref%
{FigIllust}(a1). Eq.(\ref{f2h}) is rewritten in the form of%
\begin{equation}
f_{h}^{2\text{D}}\left( \mathbf{k}\right)
=k_{x}^{5}-7k_{x}^{3}k_{y}^{2}+2k_{x}k_{y}^{4}.
\end{equation}%
However, it is impossible to implement it in the tight-binding model because
the five-fold rotational symmetry is prohibited in crystals.

\begin{figure}[t]
\centerline{\includegraphics[width=0.48\textwidth]{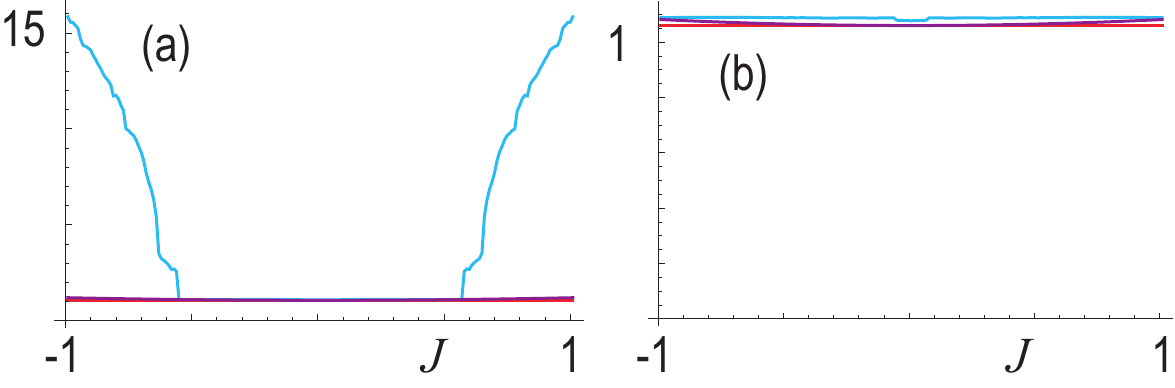}}
\caption{Fermi volume as a function of $J$. (a) $h$-wave magnets in 2D. (b) $%
j$-wave magnets in 3D. The vertical axis is the Fermi volume in units of $8%
\protect\sqrt{2}\protect\pi (m\protect\mu )^{3/2}/(3\hbar ^{3})$. The
horizontal axis is $J$ in units of $\protect\varepsilon _{0}$. Cyan curves
represent numerical results based on the tight-binding model. Red lines
represent the analytic result $V_{0}$ based on the continuum model. Purple
curves represent analytic results based on the continuum model up to the
second order in $J.$We have set $\hbar ^{2}/\left( ma^{2}\right) =\protect%
\varepsilon _{0}/4$ and $\protect\mu =0.1\protect\varepsilon _{0}$.}
\label{FigVolume}
\end{figure}

\subsection{$h$-wave magnets in 3D}

The $h$-wave magnets in 3D is described by the band-splitting function (\ref%
{f}) with the use of $f_{g}^{2\text{D}}\left( \mathbf{k}\right) $, or%
\begin{equation}
f_{h}^{3\text{D}}\left( \mathbf{k}\right) =4a^{5}k_{x}k_{y}k_{z}\left(
k_{x}^{2}-k_{y}^{2}\right) .  \label{3h}
\end{equation}%
The Fermi surface is given by%
\begin{equation}
k_{\text{F}}=\frac{\sqrt{2m\mu }}{\hbar }-sJ\frac{4\mu ^{2}m^{3}}{\hbar ^{6}}%
\cos \theta \sin 4\phi
\end{equation}%
up to the first order in $J$. Using it, the Fermi volume is given by 
\begin{equation}
V_{h}^{\text{3D}}=V_{0}^{3\text{D}}+\frac{176\sqrt{2}\pi \mu ^{9/2}m^{13/2}}{%
3\hbar ^{13}}J^{2}
\end{equation}%
with the Fermi volume in the absence of magnetic order,%
\begin{equation}
V_{0}^{3\text{D}}=\frac{8\sqrt{2}\pi }{3}\left( \frac{\mu m}{\hbar ^{2}}%
\right) ^{3/2}.  \label{V0}
\end{equation}%
In contrast to the $h$-wave magnet in 2D, it is possible to construct a
tight-binding model corresponding to the continuum model (\ref{Model}) with
Eq.(\ref{3h}) on the cubit lattice. It is given by%
\begin{align}
H=& \frac{\hbar ^{2}}{ma^{2}}\left( 3-\cos ak_{x}-\cos ak_{y}-\cos
ak_{z}\right) \sigma _{0}  \notag \\
& +2J\sigma _{z}\sin ak_{x}\sin ak_{y}\sin ak_{z}\left( \cos ak_{y}-\cos
ak_{x}\right) .
\end{align}%
The Fermi surface is shown in Fig.\ref{FigIllust}(a2) based on the
tight-binding model. The Fermi volume is shown as a function of $J$ based on
the tight-binding model without using the perturbative expansion of $J$ in
Fig.\ref{FigVolume}(a). It has almost no dependence on $J$ even for large $J$%
, where it is a good approximation to use $V_{h}^{\text{3D}}=V_{0}^{3\text{D}%
}$ with Eq.(\ref{V0}).

\subsection{$i$-wave magnets in 2D}

The continuum form of the $h$-wave magnets in 2D is given by%
\begin{equation}
f_{j}^{2\text{D}}\left( \mathbf{k}\right) =k^{7}\cos 7\phi .  \label{f2i}
\end{equation}%
The Fermi surface is given by%
\begin{equation}
k_{\text{F}}=\frac{\sqrt{2m\mu }}{\hbar }-sJ\frac{8\mu ^{3}m^{4}}{\hbar ^{8}}%
\cos 7\phi
\end{equation}%
up to the first order in $J$, and shown in Fig.\ref{FigIllust}(b1). Eq.(\ref%
{f2i}) is rewritten in the form of%
\begin{equation}
f_{j}^{2\text{D}}\left( \mathbf{k}\right)
=k_{x}^{7}-18k_{x}^{5}k_{y}^{2}+25k_{x}^{3}k_{y}^{4}-4k_{x}k_{y}^{6}.
\end{equation}%
However, it is impossible to implement it in the tight-binding model because
the seven-fold rotational symmetry is prohibited in crystals.

\begin{figure}[t]
\centerline{\includegraphics[width=0.48\textwidth]{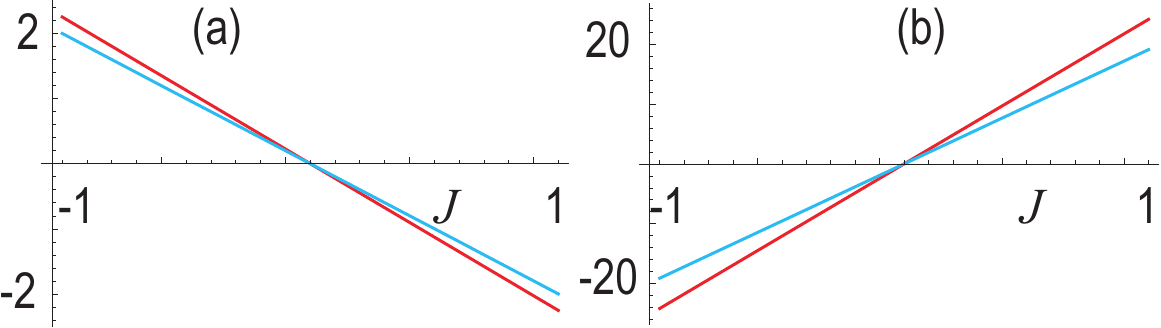}}
\caption{Spin conductivity as a function of $J$. (a) $h$-wave magnets in 2D.
We have set $\protect\mu =0.05\protect\varepsilon _{0}$. (b) $j$-wave
magnets in 3D. The vertical axis is $\protect\sigma _{\text{spin}}$ in units
of $e^{N_{X}}\protect\tau ^{N_{X}-1}\protect\varepsilon _{0}/\left( \hbar
^{N_{X}}k_{0}\right) $. The horizontal axis is $J$ in units of $\protect%
\varepsilon _{0}$. We have set $\hbar ^{2}/\left( ma^{2}\right) =\protect%
\varepsilon _{0}/4$ and $\protect\mu =0.01\protect\varepsilon _{0}.$}
\label{FigConducJ}
\end{figure}

\subsection{$j$-wave magnets in 3D}

We predict $j$-wave magnets in 3D based on the band-splitting function (\ref%
{ff}) with the use of $f_{i}^{2\text{D}}\left( \mathbf{k}\right) $, or 
\begin{equation}
f_{j}^{3\text{D}}\left( \mathbf{k}\right) =2a^{7}k_{x}k_{y}k_{z}\left(
3k_{x}^{2}-k_{y}^{2}\right) \left( k_{x}^{2}-3k_{y}^{2}\right) .  \label{3j}
\end{equation}%
The Fermi surface is given by%
\begin{equation}
k_{\text{F}}=\frac{\sqrt{2m\mu }}{\hbar }-sJ\frac{8\mu ^{3}m^{4}}{\hbar ^{8}}%
\cos \theta \sin 6\phi
\end{equation}%
up to the first order in $J$. Using it, the Fermi volume is given by%
\begin{equation}
V_{j}^{3\text{D}}=V_{0}^{3\text{D}}+\frac{128\sqrt{2}\pi \mu ^{13/2}m^{17/2}%
}{3\hbar ^{17}}J^{2}.
\end{equation}%
In contrast to the $i$-wave magnet in 2D, it is possible to construct a
tight-binding model corresponding to the continuum model (\ref{Model}) with
Eq.(\ref{3j}) on the triangular prism lattice. It is given by%
\begin{align}
H=& \Bigg[\frac{-2\hbar ^{2}}{3ma^{2}}\Big(\sum_{j=0}^{2}\cos (a\mathbf{n}%
_{j}^{\text{A}}\cdot \mathbf{k})-3\Big)  \notag \\
& \hspace{20mm}+\frac{\hbar ^{2}}{m_{z}a^{2}}(1-\cos ak_{z})\Bigg]\sigma _{0}
\notag \\
& +16J\sigma _{z}\sin ak_{z}\prod\limits_{j=0}^{2}\sin (a\mathbf{n}_{j}^{%
\text{A}}\cdot \mathbf{k})\prod\limits_{j=0}^{2}\sin (\sqrt{3}a\mathbf{n}%
_{j}^{\text{B}}\cdot \mathbf{k}),
\end{align}%
where we have defined $\mathbf{n}_{j}^{\text{A}}=\left( \cos \frac{2\pi j}{3}%
,\sin \frac{2\pi j}{3}\right) $ and $\mathbf{n}_{j}^{\text{B}}=\left( \sin 
\frac{2\pi j}{3},\cos \frac{2\pi j}{3}\right) $ with $j=0,1,2$.

The Fermi surface is shown in Fig.\ref{FigIllust}(b2) based on the
tight-binding model. The Fermi volume is shown as a function of $J$ based on
the tight-binding model without using the perturbative expansion of $J$ in
Fig.\ref{FigVolume}(b). It has almost no dependence on $J$ even for large $J$%
, where it is a good approximation to use $V_{j}^{\text{3D}}=V_{0}^{3\text{D}%
}$.

\begin{figure}[t]
\centerline{\includegraphics[width=0.48\textwidth]{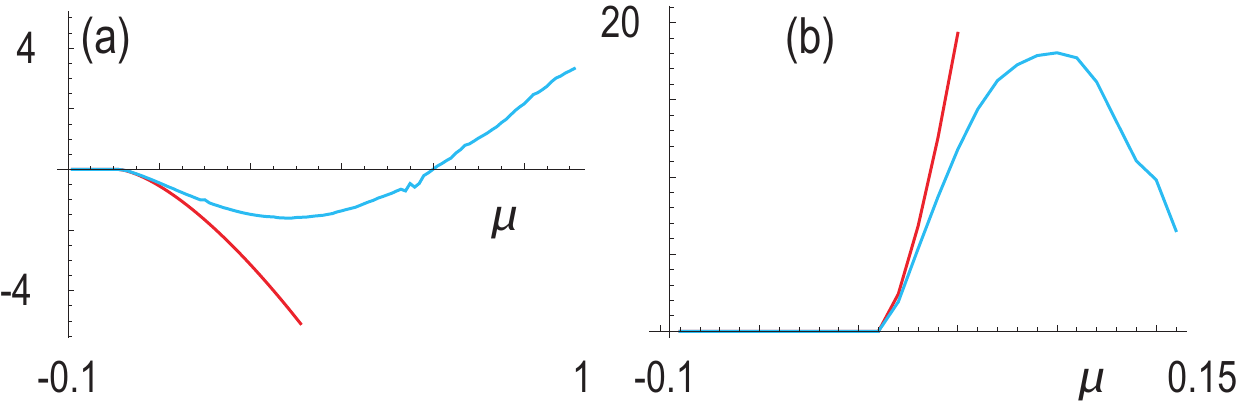}}
\caption{Spin conductivity as a function of $\protect\mu $. (a) $h$-wave
magnets in 2D. (b) $j$-wave magnets in 3D. The vertical axis is $\protect%
\sigma _{\text{spin}}$ in units of $e^{N_{X}}\protect\tau ^{N_{X}-1}\protect%
\varepsilon _{0}/\left( \hbar ^{N_{X}}k_{0}\right) $. The horizontal axis is 
$\protect\mu $ in units of $\protect\varepsilon _{0}$. We have set $\hbar
^{2}/\left( ma^{2}\right) =\protect\varepsilon _{0}/4$ and $J=0.1\protect%
\varepsilon _{0}$.}
\label{FigConducMu}
\end{figure}

\section{Nonlinear spin current generation}

$X$-wave magnets have a characteristic nonlinear spin conductivity, where
only a certain order of nonlinear spin conductivity is nonzero\cite{GI}. For
example, only the second-order nonlinear spin conductivity is nonzero for $f$%
-wave magnets\cite{GI}.

The formula describing the nonlinear spin current by the application of $%
E_{x}$ and $E_{y}$ has been studied\cite{GI}. It is generalized to%
\begin{equation}
j_{b}=\sum_{\ell _{x},\ell _{y},\ell _{z}}\sigma ^{x^{\ell _{x}}y^{\ell
_{y}}z^{\ell _{z}};b}(E_{x})^{\ell _{x}}(E_{y})^{\ell _{y}}(E_{z})^{\ell
_{z}}  \label{Expansion}
\end{equation}%
by the application of $E_{x}$, $E_{y}$\ and $E_{z}$. Then, the $(\ell
_{x}+\ell _{y}+\ell _{z})$-th order conductivity is defined by

\begin{equation}
\sigma ^{x^{\ell _{x}}y^{\ell _{y}}z^{\ell _{z}};b}=\frac{1}{\ell _{x}!\ell
_{y}!\ell _{z}!}\frac{\partial ^{\ell _{x}+\ell _{y}+\ell _{z}}j_{b}}{%
\partial E_{x}^{\ell _{x}}\partial E_{y}^{\ell _{y}}\partial E_{z}^{\ell
_{z}}}.
\end{equation}%
The spin-dependent conductivity is explicitly given by%
\begin{eqnarray}
&&\sigma _{s}^{x^{\ell _{x}}y^{\ell _{y}}z^{\ell _{z}};b}  \notag \\
&=&\frac{\left( -\frac{e}{\hbar }\right) ^{\ell _{x}+\ell _{y}+\ell _{z}+1}}{%
\left( i\omega +\frac{1}{\tau }\right) ^{\ell _{x}+\ell _{y}+\ell _{z}}}\int
d\mathbf{k}\,f_{s}^{\left( 0\right) }\frac{\partial ^{\ell _{x}+\ell
_{y}+\ell _{z}+1}\varepsilon _{s}}{\partial k_{x}^{\ell _{x}}\partial
k_{y}^{\ell _{y}}\partial k_{z}^{\ell _{z}}\partial k_{b}}.  \label{Drude}
\end{eqnarray}%
Based on this formula, we find that the spin conductivity is defined%
\begin{equation}
\sigma _{\text{spin}}^{x^{\ell _{x}}y^{\ell _{y}}z^{\ell _{z}+1};b}=\frac{%
\sigma _{\uparrow }^{x^{\ell _{x}}y^{\ell _{y}}z^{\ell _{z}+1};b}-\sigma
_{\downarrow }^{x^{\ell _{x}}y^{\ell _{y}}z^{\ell _{z}+1};b}}{2}
\label{spinDr}
\end{equation}%
for $\left( X+1\right) $-wave odd-parity magnets in 3D is identical to $%
\sigma _{\text{spin}}^{x^{\ell _{x}}y^{\ell _{y}}z^{\ell _{z}};b}$ for $X$%
-wave altermagnets in 2D.

The energy of the $X$-wave magnets is expanded in the form of%
\begin{equation}
\varepsilon _{s}=\frac{\hbar ^{2}k^{2}}{2m}+sJ%
\sum_{n_{x}+n_{y}+n_{z}=N_{X}}c^{n_{x},n_{y},n_{z}}k_{x}^{n_{x}}k_{y}^{n_{y}}k_{z}^{n_{z}}.
\label{ek}
\end{equation}%
Then, the\ spin conductivity $\sigma _{\text{spin}}^{x^{\ell _{x}}y^{\ell
_{y}}z^{\ell _{z}};b}$ is nonzero only for nonzero $c^{n_{x},n_{y},n_{z}}$,
where the spin conductivity is given by%
\begin{align}
& \sigma _{\text{spin}}^{x^{n_{x}-1}y^{n_{y}}z^{n_{z}};x}  \notag \\
=& \sigma _{\text{spin}}^{x^{n_{x}}y^{n_{y}-1}z^{n_{z}};y}=\sigma _{\text{%
spin}}^{x^{n_{x}}y^{n_{y}}z^{n_{z}-1};z}  \notag \\
=& \frac{\left( -\frac{e}{\hbar }\right) ^{n_{x}+n_{y}+n_{z}+1}}{\left(
i\omega +\frac{1}{\tau }\right) ^{n_{x}+n_{y}+n_{z}}}\,V\frac{%
n_{x}!n_{y}!n_{z}!}{\left( n_{x}-\ell _{x}\right) !\left( n_{y}-\ell
_{y}\right) !\left( n_{z}-\ell _{z}\right) !}
\end{align}%
with the Fermi volume $V\equiv \int d\mathbf{k}\,f_{s}^{\left( 0\right) }$.
Hence, the function $f_{X}\left( \mathbf{k}\right) $ of the $X$-wave magnet
is experimentally determined by measuring $\sigma _{\text{spin}%
}^{x^{n_{x}}y^{n_{y}}z^{n_{z}};b}$ for various $n_{x},n_{y},n_{z}$
satisfying $n_{x}+n_{y}+n_{z}=N_{X}$. Namely, the $X$-wave magnet is
completely identified by measuring the nonlinear spin conductivities.

\subsection{$h$-wave magnets in 3D}

Using Eq.(\ref{Drude}) and the expansion Eq.(\ref{ek}), all of the nonzero
contributions are listed as $\sigma _{\text{spin}}^{x^{3}y;z}$, $\sigma _{%
\text{spin}}^{xy^{3};z}$, $\sigma _{\text{spin}}^{x^{2}yz;x}$, $\sigma _{%
\text{spin}}^{y^{3}z;x}$, $\sigma _{\text{spin}}^{x^{3}z;y}$ and $\sigma _{%
\text{spin}}^{xy^{2}z;y}$, where%
\begin{align}
\sigma _{\text{spin}}^{x^{3}y;z}& =\sigma _{\text{spin}}^{x^{2}yz;x}=\sigma
_{\text{spin}}^{x^{3}z;y}=-\sigma _{\text{spin}}^{xy^{3};z}  \notag \\
& =-\sigma _{\text{spin}}^{y^{3}z;x}=-\sigma _{\text{spin}}^{xy^{2}z;y}=6%
\frac{\left( -e/\hbar \right) ^{5}}{\left( i\omega +1/\tau \right) ^{4}}V.
\end{align}%
The spin conductivity is shown as a function of $J$ in Fig.\ref{FigConducJ}%
(a). It is linear even for the numerical calculation without using the
perturbation theory with respect to $J$. The spin conductivity is shown as a
function of $\mu $ in Fig.\ref{FigConducMu}(a). The analytic formula based
on the continuum model well reproduces the numerical result based on the
tight-binding model at the bottom of the band structure.

\subsection{$j$-wave magnets in 3D}

Using Eq.(\ref{Drude}) and the expansion Eq.(\ref{ek}), all of the nonzero
contributions are $\sigma _{\text{spin}}^{x^{5}y;z}$, $\sigma _{\text{spin}%
}^{x^{3}y^{3};z}$, $\sigma _{\text{spin}}^{xy^{5};z}$, $\sigma _{\text{spin}%
}^{x^{4}yz;x}$, $\sigma _{\text{spin}}^{x^{2}y^{3}z;x}$, $\sigma _{\text{spin%
}}^{y^{5}z;x}$, $\sigma _{\text{spin}}^{x^{5}z;y}$, $\sigma _{\text{spin}%
}^{x^{3}y^{2}z;y}$ and $\sigma _{\text{spin}}^{xy^{4}z;y}$, where%
\begin{align}
\sigma _{\text{spin}}^{x^{5}y;z}& =\sigma _{\text{spin}}^{x^{4}yz;x}=\sigma
_{\text{spin}}^{x^{5}z;y}=\sigma _{\text{spin}}^{xy^{5};z}=\sigma _{\text{%
spin}}^{y^{5}z;x}  \notag \\
& =\sigma _{\text{spin}}^{xy^{4}z;y}=-\sigma _{\text{spin}%
}^{x^{3}y^{3};z}=-\sigma _{\text{spin}}^{x^{2}y^{3}z;x}=-\sigma _{\text{spin}%
}^{x^{3}y^{2}z;y}  \notag \\
& =6!\frac{\left( -e/\hbar \right) ^{7}}{\left( i\omega +1/\tau \right) ^{6}}%
V.
\end{align}%
The spin conductivity is shown as a function of $J$ in Fig.\ref{FigConducJ}%
(b). It is linear even for the numerical calculation without using the
perturbation theory with respect to $J$. The spin conductivity is shown as a
function of $\mu $ in Fig.\ref{FigConducMu}(b). The analytic formula based
on the continuum model well reproduces the numerical result based on the
tight-binding model at the bottom of the band structure.

\section{Discussions}

We have proposed a scheme to construct $X$-wave magnets in 3D by a
dimensional extension of $X$-wave magnets in 2D. Then, $h$-wave magnets are
the 3D extension of $g$-wave altermagnets, while $j$-wave magnets are the 3D
extension of $i$-wave altermagnets. They can be identified experimentally by
measuring nonlinear spin currents by the application of an electric field.
Only the fourth-order (sixth-order) spin currents flow in $h$-wave ($j$%
-wave) magnets, which function as spin-current  rectifier  because the
spin current exhibits unidirectional flow independent of the applied
electric field.

A group-theoretical framework for identifying odd-parity magnets has
recently been proposed\cite{YuH}. Using this framework, $h$-wave spin
splitting was theoretically predicted in FeSe through density functional
theory\cite{YuH}. In a similar way, $j$-wave magnets are expected to be
discovered in the near future. Both $h$-wave and $j$-wave odd-parity magnets
in three dimensions have been theoretically proposed using the
layer-construction method, where candidate materials include layered $g$%
-wave and $i$-wave altermagnets. It is known that $p$-wave magnets arise
from a helical spin structure, in which a linear $k_{z}$ term appears when
the helix rotates along the $z$ direction. By analogy, we expect that $h$%
-wave and $j$-wave odd-parity magnets can emerge from layered structures
composed of $g$-wave and $i $-wave altermagnets, where the spins rotate
along the $z$ direction. Two-dimensional $g$-wave and $i$-wave altermagnets
have been theoretically proposed in twisted magnetic van der Waals bilayers%
\cite{YLiu}, and $i$-wave altermagnet has also been proposed in MnP(S,Se)$%
_{3}$\cite{MazinIwave}. Therefore, stacking such systems may provide
promising candidates for realizing $h$-wave and $j$-wave odd-parity magnets.

We estimate the magnitude of electric field required to generate nonlinear
spin currents\cite{GI}. The spin conductivity with the dimensionless
electric field is $E/E_{\text{cr}}$ with the critical electric field $E_{%
\text{cr}}\equiv \frac{\hbar k}{e\tau }=0.66\left[ \text{V}/\mu \text{m}%
\right] $. Here we have used the relaxation time\cite{Trama,Du} $\tau
=3.4\times 10^{-12}$s and the typical momentum\cite{Krem} $k=0.35$\AA $^{-1} 
$. A typical experimental electric field\cite{Sala,PHe} is $E=1$[V/$\mu $m],
which satisfies the condition $E/E_{\text{cr}}>1$. Therefore, higher-order
nonlinear spin currents can be generated under realistic experimental
conditions.

A strong electric field is required to induce magnetization, which may lead
to Joule-heating issues. This problem can be mitigated by applying pulsed
electric fields. In addition, cooling with a refrigerator is effective for
suppressing Joule heating.

This work is supported by CREST, JST (Grants No. JPMJCR20T2) and
Grants-in-Aid for Scientific Research from MEXT KAKENHI (Grant No. 23H00171).%

\appendix

\beginWide

\section{Spherical harmonic functions}

For the sake of the completeness, we classify all X-wave magnets in the
following tables. They are classification tables for $X$-wave magnets in
terms of the point group (PG), the magnetic point group (MPG) and the
irreducible representation (IR). "NO" in the tables given by Eqs.(A8) and
(A10) represents that the spherical harmonic function is not compatible with
the lattice structure due to the five-fold and seven-fold rotational
symmetries.

\subsection{$p$-wave}

\begin{equation}
\begin{array}{|c|c|c|c|c|}
\hline
p\text{-waves} &  & \text{Spherical harmonic functions} & \text{PG (MPG)} & 
\text{IR} \\ \hline
& Q_{10}^{-} & 0 &  &  \\ \hline
& Q_{10}^{+} & k_{z} & D_{2h}(mmm^{\prime }) & B_{1u} \\ \hline
f_{p^{\prime }}^{2\text{D}} & Q_{11}^{-} & k_{y} & D_{2h}(mm^{\prime }m) & 
B_{2u} \\ \hline
f_{p}^{2\text{D}} & Q_{11}^{+} & k_{x} & D_{2h}(m^{\prime }mm) & B_{3u} \\ 
\hline
\end{array}%
\end{equation}

\subsection{$d$-wave}

\begin{equation}
\begin{array}{|c|c|c|c|c|}
\hline
d\text{-waves} &  & \text{Spherical harmonic functions} & \text{PG (MPG)} & 
\text{IR} \\ \hline
& Q_{20}^{-} & 0 &  &  \\ \hline
& Q_{20}^{+} & 3k_{z}^{2}-k^{2} & \left( mmm\right) & A_{1g} \\ \hline
& Q_{21}^{-} & k_{y}k_{z} & D_{2h}(mm^{\prime }m^{\prime }) & E_{g} \\ \hline
& Q_{21}^{+} & k_{x}k_{z} & D_{2h}(m^{\prime }mm^{\prime }) & E_{g} \\ \hline
f_{d}^{2\text{D}} & Q_{22}^{-} & k_{x}k_{y} & D_{4h}(4^{\prime }/m^{\prime
}m^{\prime }m) & B_{2g} \\ \hline
f_{d^{\prime }}^{2\text{D}} & Q_{22}^{+} & k_{x}^{2}-k_{y}^{2} & 
D_{4h}(4^{\prime }mmm) & B_{1g} \\ \hline
\end{array}
\label{Tab-d}
\end{equation}

There are relations 
\begin{equation}
Q_{21}^{+}\propto k_{z}Q_{11}^{+},\qquad Q_{21}^{-}\propto k_{z}Q_{11}^{-}.
\end{equation}

\subsection{$f$-wave}

\begin{equation}
\begin{array}{|c|c|c|c|c|}
\hline
f\text{-waves} &  & \text{Spherical harmonic functions} & \text{PG (MPG)} & 
\text{IR} \\ \hline
& Q_{30}^{-} & 0 &  &  \\ \hline
& Q_{30}^{+} & k_{z}\left( 5k_{z}^{2}-3k^{2}\right)  & \left( mmm^{\prime
}\right)  & A_{1g} \\ \hline
& Q_{31}^{-} & k_{y}\left( k^{2}-5k_{z}^{2}\right)  & D_{2h}(mm^{\prime }m)
& B_{2u} \\ \hline
& Q_{31}^{+} & k_{x}\left( k^{2}-5k_{z}^{2}\right)  & D_{2h}(m^{\prime }mm)
& B_{3u} \\ \hline
f_{f}^{3\text{D}} & Q_{32}^{-} & 2k_{x}k_{y}k_{z} & O_{h}(4^{\prime
}/m^{\prime }m^{\prime }m^{\prime }) & A_{2u} \\ \hline
f_{f^{\prime }}^{3\text{D}} & Q_{32}^{+} & \left( k_{x}^{2}-k_{y}^{2}\right)
k_{z} & D_{4h}(4^{\prime }/mmm^{\prime }) & B_{1g} \\ \hline
f_{f^{\prime }}^{2\text{D}} & Q_{33}^{-} & k_{y}\left(
3k_{x}^{2}-k_{y}^{2}\right)  & D_{6h}(6^{\prime }/mm^{\prime }m) & E_{2u} \\ 
\hline
f_{f}^{2\text{D}} & Q_{33}^{+} & k_{x}\left( k_{x}^{2}-3k_{y}^{2}\right)  & 
D_{6h}(6^{\prime }/m^{\prime }mm) & E_{2u} \\ \hline
\end{array}
\label{Tab-f}
\end{equation}

There are relations 
\begin{eqnarray}
Q_{32}^{+} \propto k_{z}Q_{22}^{+}, \qquad Q_{32}^{-} \propto
k_{z}Q_{22}^{-}.
\end{eqnarray}

\subsection{$g$-wave}

\begin{equation}
\begin{array}{|c|c|c|c|c|}
\hline
g\text{-waves} &  & \text{Spherical harmonic functions} & \text{PG (MPG)} & 
\text{IR} \\ \hline
& Q_{40}^{-} & 0 &  &  \\ \hline
& Q_{40}^{+} & 35k_{z}^{4}-30k_{z}^{2}k^{2}+3k^{4} & \left( mmm\right)  & 
A_{1g} \\ \hline
& Q_{41}^{-} & k_{z}k_{y}\left( 7k_{z}^{2}-3k^{2}\right)  & 
D_{2h}(mm^{\prime }m) & B_{2u} \\ \hline
& Q_{41}^{+} & k_{z}k_{x}\left( 7k_{z}^{2}-3k^{2}\right)  & D_{2h}(m^{\prime
}mm) & B_{3u} \\ \hline
& Q_{42}^{-} & 2k_{x}k_{y}\left( 7k_{z}^{2}-k^{2}\right)  & D_{4h}(4^{\prime
}/m^{\prime }m^{\prime }m) & B_{2g} \\ \hline
& Q_{42}^{+} & \left( k_{x}^{2}-k_{y}^{2}\right) \left(
7k_{z}^{2}-k^{2}\right)  & D_{4h}(4^{\prime }/mmm) & B_{1g} \\ \hline
f_{g^{\prime }}^{3\text{D}} & Q_{43}^{-} & k_{z}k_{y}\left(
3k_{x}^{2}-k_{y}^{2}\right)  & D_{6h}(6^{\prime }/mm^{\prime }m^{\prime }) & 
E_{2g} \\ \hline
f_{g}^{3\text{D}} & Q_{43}^{+} & k_{z}k_{x}\left(
k_{x}^{2}-3k_{y}^{2}\right)  & D_{6h}(6^{\prime }/m^{\prime }mm^{\prime }) & 
E_{2g} \\ \hline
f_{g}^{2\text{D}} & Q_{44}^{-} & k_{x}k_{y}\left( k_{x}^{2}-k_{y}^{2}\right) 
& D_{4h}(4/m^{\prime }m^{\prime }m) & A_{2g} \\ \hline
f_{g^{\prime }}^{2\text{D}} & Q_{44}^{+} & 
k_{x}^{4}-6k_{x}^{2}k_{y}^{2}+k_{y}^{4} & D_{4h}(4/mmm) & A_{1g} \\ \hline
\end{array}
\label{Tab-g}
\end{equation}

There are relations 
\begin{equation}
Q_{43}^{+}\propto k_{z}Q_{33}^{+},\qquad Q_{43}^{-}\propto k_{z}Q_{33}^{-}.
\end{equation}

\subsection{$h$-wave}

\begin{equation}
\begin{array}{|c|c|c|c|c|}
\hline
h\text{-waves} &  & \text{Spherical harmonic functions} & \text{PG (MPG)} & 
\text{IR} \\ \hline
& Q_{50}^{-} & 0 &  &  \\ \hline
& Q_{50}^{+} & k_{z}\left( 63k_{z}^{4}-70k_{z}^{2}k^{2}+15k^{4}\right)  & 
\left( mmm^{\prime }\right)  & A_{1g} \\ \hline
& Q_{51}^{-} & k_{y}\left( 21k_{z}^{4}-14k_{z}^{2}k^{2}+k^{4}\right)  & 
D_{2h}(mm^{\prime }m) & B_{2u} \\ \hline
& Q_{51}^{+} & k_{x}\left( 21k_{z}^{4}-14k_{z}^{2}k^{2}+k^{4}\right)  & 
D_{2h}(m^{\prime }mm) & B_{3u} \\ \hline
& Q_{52}^{-} & 2k_{z}k_{x}k_{y}\left( k_{z}^{2}-k^{2}\right)  & 
D_{4h}(4^{\prime }/m^{\prime }m^{\prime }m^{\prime }) & B_{2g} \\ \hline
& Q_{52}^{+} & k_{z}\left( k_{x}^{2}-k_{y}^{2}\right) \left(
k_{z}^{2}-k^{2}\right)  & D_{4h}(4^{\prime }/mmm^{\prime }) & B_{1g} \\ 
\hline
& Q_{53}^{-} & k_{y}\left( 3k_{x}^{2}-k_{y}^{2}\right) \left(
9k_{z}^{2}-k^{2}\right)  & D_{6h}(6^{\prime }/mm^{\prime }m) & E_{2u} \\ 
\hline
& Q_{53}^{+} & k_{x}\left( k_{x}^{2}-3k_{y}^{2}\right) \left(
9k_{z}^{2}-k^{2}\right)  & D_{6h}(6^{\prime }/m^{\prime }mm) & E_{2u} \\ 
\hline
f_{h}^{3\text{D}} & Q_{54}^{-} & k_{z}k_{x}k_{y}\left(
k_{x}^{2}-k_{y}^{2}\right)  & D_{4h}(4/m^{\prime }m^{\prime }m^{\prime }) & 
A_{1u} \\ \hline
f_{h^{\prime }}^{3\text{D}} & Q_{54}^{+} & k_{z}\left(
k_{x}^{4}-6k_{x}^{2}k_{y}^{2}+k_{y}^{4}\right)  & D_{4h}(4/mmm^{\prime }) & 
A_{2u} \\ \hline
\text{NO} & Q_{55}^{-} & 5k_{x}^{4}k_{y}-10k_{x}^{2}k_{y}^{3}+k_{y}^{5} & 
\text{NO} &  \\ \hline
\text{NO} & Q_{55}^{+} & k_{x}^{5}-10k_{x}^{3}k_{y}^{2}+5k_{x}k_{y}^{4} & 
\text{NO} &  \\ \hline
\end{array}
\label{Tab-h}
\end{equation}

There are relations 
\begin{equation}
Q_{54}^{+}\propto k_{z}Q_{44}^{+},\qquad Q_{54}^{-}\propto k_{z}Q_{44}^{-}.
\end{equation}

\subsection{$i$-wave}

\begin{equation}
\begin{array}{|c|c|c|c|c|}
\hline
i\text{-waves} &  & \text{Spherical harmonic functions} & \text{PG (MPG)} & 
\text{IR} \\ \hline
& Q_{60}^{-} & 0 &  &  \\ \hline
& Q_{60}^{+} & 35k_{z}^{6}-105k_{z}^{4}k^{2}+63k_{z}^{2}k^{4}-5k^{6} & (mmm)
& A_{1g} \\ \hline
& Q_{61}^{-} & k_{z}k_{y}\left( 33k_{z}^{4}-30k_{z}^{2}k^{2}+5k^{4}\right) 
& D_{2h}(mm^{\prime }m) & B_{2u} \\ \hline
& Q_{61}^{+} & k_{z}k_{x}\left( 33k_{z}^{4}-30k_{z}^{2}k^{2}+5k^{4}\right) 
& D_{2h}(m^{\prime }mm) & B_{3u} \\ \hline
& Q_{62}^{-} & 2k_{x}k_{y}\left( 33k_{z}^{4}-18k_{z}^{2}k^{2}+k^{4}\right) 
& D_{4h}(4^{\prime }/m^{\prime }m^{\prime }m^{\prime }) & B_{2g} \\ \hline
& Q_{62}^{+} & \left( k_{x}^{2}-k_{y}^{2}\right) \left(
33k_{z}^{4}-18k_{z}^{2}k^{2}+k^{4}\right)  & D_{4h}(4^{\prime }/mmm^{\prime
}) & B_{1g} \\ \hline
& Q_{63}^{-} & k_{z}k_{y}\left( 3k_{x}^{2}-k_{y}^{2}\right) \left(
11k_{z}^{2}-3k^{2}\right)  & D_{6h}(6^{\prime }/mm^{\prime }m^{\prime }) & 
E_{2u} \\ \hline
& Q_{63}^{+} & k_{z}k_{x}\left( k_{x}^{2}-3k_{y}^{2}\right) \left(
11k_{z}^{2}-3k^{2}\right)  & D_{6h}(6^{\prime }/m^{\prime }mm^{\prime }) & 
E_{2u} \\ \hline
& Q_{64}^{-} & 4k_{x}k_{y}\left( k_{x}^{2}-k_{y}^{2}\right) \left(
11k_{z}^{2}-k^{2}\right)  & D_{4h}(4/m^{\prime }m^{\prime }m) & A_{2g} \\ 
\hline
& Q_{64}^{+} & \left( k_{x}^{4}-6k_{x}^{2}k_{y}^{2}+k_{y}^{4}\right) \left(
11k_{z}^{2}-k^{2}\right)  & D_{4h}(4/mmm) & A_{1g} \\ \hline
\text{NO} & Q_{65}^{-} & k_{z}\left(
5k_{x}^{4}k_{y}-10k_{x}^{2}k_{y}^{3}+k_{y}^{5}\right)  & \text{NO} & E_{g}
\\ \hline
\text{NO} & Q_{65}^{+} & k_{z}\left(
k_{x}^{5}-10k_{x}^{3}k_{y}^{2}+5k_{x}k_{y}^{4}\right)  & \text{NO} & E_{g}
\\ \hline
f_{i}^{2\text{D}} & Q_{66}^{-} & 2k_{x}k_{y}\left(
3k_{x}^{2}-k_{y}^{2}\right) \left( k_{x}^{2}-3k_{y}^{2}\right)  & 
D_{6h}(6/m^{\prime }m^{\prime }m) & A_{2g} \\ \hline
f_{i^{\prime }}^{2\text{D}} & Q_{66}^{+} & \left( k_{x}^{2}-k_{y}^{2}\right)
\left( k_{x}^{4}+k_{y}^{4}-14k_{x}^{2}k_{y}^{2}\right)  & D_{6h}(6/mmm) & 
A_{1g} \\ \hline
\end{array}
\label{Tab-i}
\end{equation}

There are relations 
\begin{equation}
Q_{65}^{+}\propto k_{z}Q_{55}^{+},\qquad Q_{65}^{-}\propto k_{z}Q_{55}^{-}.
\end{equation}%
A comment is in order with respect to $f_{i}^{3\text{D}}\left( \mathbf{k}%
\right) $ in Eq.(\ref{EqY}), which is not included in the table (A10). It is
explicitly given by 
\begin{equation}
f_{i}^{3\text{D}}\left( \mathbf{k}\right) \propto Q_{66}^{+}-Q_{62}^{+}.
\end{equation}

\subsection{$j$-wave}

\begin{equation}
\begin{array}{|c|c|c|c|c|}
\hline
j\text{-waves} &  & \text{Spherical harmonic functions} & \text{PG (MPG)} & 
\text{IR} \\ \hline
& Q_{70}^{-} & 0 &  &  \\ \hline
& Q_{70}^{+} & k_{z}\left(
429k_{z}^{6}-693k_{z}^{4}k^{2}+315k_{z}^{2}k^{4}-35k^{6}\right)  & 
(mmm^{\prime }) & A_{1g} \\ \hline
& Q_{71}^{-} & k_{y}\left(
429k_{z}^{6}-495k_{z}^{4}k^{2}+135k_{z}^{2}k^{4}-5k^{6}\right)  & 
D_{2h}(mm^{\prime }m) & B_{2u} \\ \hline
& Q_{71}^{+} & k_{x}\left(
429k_{z}^{6}-495k_{z}^{4}k^{2}+135k_{z}^{2}k^{4}-5k^{6}\right)  & 
D_{2h}(m^{\prime }mm) & B_{3u} \\ \hline
& Q_{72}^{-} & k_{z}k_{x}k_{y}\left(
143k_{z}^{4}-110k_{z}^{2}k^{2}+15k^{4}\right)  & D_{4h}(4^{\prime }m^{\prime
}m^{\prime }m^{\prime }) & B_{2g} \\ \hline
& Q_{72}^{+} & k_{z}\left( k_{x}^{2}-k_{y}^{2}\right) \left(
143k_{z}^{4}-110k_{z}^{2}k^{2}+15k^{4}\right)  & D_{4h}(4^{\prime
}/mmm^{\prime }) & B_{1g} \\ \hline
& Q_{73}^{-} & k_{y}\left( 3k_{x}^{2}-k_{y}^{2}\right) \left(
143k_{z}^{4}-66k_{z}^{2}k^{2}+3k^{4}\right)  & D_{6h}(6^{\prime }mm^{\prime
}m) & E_{2u} \\ \hline
& Q_{73}^{+} & k_{x}\left( k_{x}^{2}-3k_{y}^{2}\right) \left(
143k_{z}^{4}-66k_{z}^{2}k^{2}+3k^{4}\right)  & D_{6h}(6^{\prime }mm^{\prime
}m) & E_{2u} \\ \hline
& Q_{74}^{-} & k_{z}k_{x}k_{y}\left( k_{x}^{2}-k_{y}^{2}\right) \left(
13k_{z}^{2}-3k^{2}\right)  & D_{4h}(4/m^{\prime }m^{\prime }m^{\prime }) & 
A_{2g} \\ \hline
& Q_{74}^{+} & k_{z}\left( k_{x}^{4}-6k_{x}^{2}k_{y}^{2}+k_{y}^{4}\right)
\left( 13k_{z}^{2}-3k^{2}\right)  & D_{4h}(4/mmm^{\prime }) & A_{1g} \\ 
\hline
\text{NO} & Q_{75}^{-} & \left(
5k_{x}^{4}k_{y}-10k_{x}^{2}k_{y}^{3}+k_{y}^{5}\right) \left(
13k_{z}^{2}-k^{2}\right)  & \text{NO} & E_{1u} \\ \hline
\text{NO} & Q_{75}^{+} & \left(
k_{x}^{5}-10k_{x}^{3}k_{y}^{2}+5k_{x}k_{y}^{4}\right) \left(
13k_{z}^{2}-k^{2}\right)  & \text{NO} & E_{1u} \\ \hline
f_{j}^{3\text{D}} & Q_{76}^{-} & 2k_{z}k_{x}k_{y}\left(
3k_{x}^{2}-k_{y}^{2}\right) \left( k_{x}^{2}-3k_{y}^{2}\right)  & 
D_{6h}(6/m^{\prime }m^{\prime }m^{\prime }) & A_{2u} \\ \hline
f_{j^{\prime }}^{3\text{D}} & Q_{76}^{+} & k_{z}\left(
k_{x}^{2}-k_{y}^{2}\right) \left(
k_{x}^{4}+k_{y}^{4}-14k_{x}^{2}k_{y}^{2}\right)  & D_{6h}(6/mmm^{\prime }) & 
A_{1u} \\ \hline
\text{NO} & Q_{77}^{-} & 
7k_{x}^{6}k_{y}-35k_{x}^{4}k_{y}^{3}+21k_{x}^{2}k_{y}^{5}-7k_{x}k_{y}^{6} & 
\text{NO} & E_{1u} \\ \hline
\text{NO} & Q_{77}^{+} & 
k_{x}^{7}-21k_{x}^{5}k_{y}^{2}+35k_{x}^{3}k_{y}^{4}-7k_{x}k_{y}^{6} & \text{%
NO} & E_{1u} \\ \hline
\end{array}
\label{Tab-j}
\end{equation}

There are relations 
\begin{equation}
Q_{76}^{+}\propto k_{z}Q_{66}^{+},\qquad Q_{76}^{-}\propto k_{z}Q_{66}^{-}.
\end{equation}

\endWide

\end{document}